# Rapid and robust whole slide imaging based on LED-array illumination and color-multiplexed single-shot autofocusing

Shaowei Jiang[1,4], Zichao Bian[1,4], Xizhi Huang[1,3,4], Pengming Song[2], He Zhang[1], Yongbing Zhang[3], and Guoan Zheng[1,2]

[1]*Department of Biomedical Engineering, University of Connecticut, Storrs, 06269, USA*

[2]*Department of Electrical and Computer Engineering, University of Connecticut, Storrs, 06269, USA*

[3]*Graduate School at Shenzhen, Tsinghua University, Shenzhen, 518055, China*

[4]*These authors contributed equally to this work*

*Correspondence to: Guoan Zheng, Department of Biomedical Engineering, University of Connecticut, Storrs, 06269, USA*
*Email: guoan.zheng@uconn.edu*

**Background**: Digital pathology is experiencing an exponential period of growth catalyzed by advancements in imaging hardware and progresses in machine learning. The use of whole slide imaging (WSI) for digital pathology has recently been cleared for primary diagnosis in the US. The demand for using frozen section procedure for rapid identification of cancerous tissue during surgery is another driving force for the development of WSI. A conventional WSI system scans the tissue slide to different positions and acquires the digital images. In a typical implementation, a focus map is created prior to the scanning process, leading to significant overhead time and a necessity for high positional accuracy of the mechanical system. The resulting cost of WSI system is often prohibitive for frozen section procedure during surgery.

**Methods**: We report a novel WSI scheme based on a programmable LED array for sample illumination. In between two regular brightfield image acquisitions, we acquire one additional image by turning on a red and a green LED for color multiplexed illumination. We then identify the translational shift of the red- and green-channel images by maximizing the image mutual information or cross-correlation. The resulting translational shift is used for dynamic focus correction in the scanning process. Since we track the differential focus during adjacent acquisitions, there is no positional repeatability requirement in our scheme.

**Results**: We demonstrate a prototype WSI platform with a mean focusing error of ~0.3 microns. Different from previous implementations, this prototype platform requires no focus map surveying, no secondary camera or additional optics, and allows for continuous sample motion in the focus tracking process.

**Conclusions**: A programmable LED array can be used for color-multiplexed single-shot autofocusing in WSI. The reported scheme may enable the development of cost-effective WSI platforms without positional repeatability requirement. It may also provide a turnkey solution for other high-content microscopy applications.

**Introduction**

Whole slide imaging (WSI), which refers to the scanning of a conventional tissue slide in order to produce a digital representation, promises better and faster predication, diagnosis, and prognosis of cancers and other diseases (1). Digital pathology based on WSI is now experiencing an exponential period of growth catalyzed by advancements in imaging hardware and progresses in machine learning. The regulatory field for digital pathology has advanced significantly in the past years as well. It was recommended that manufacturers of WSI devices for primary diagnosis submit applications to the US Food and Drug Administration (FDA) through the de novo process (2). A major milestone was accomplished in 2017 when the FDA approved Philips' WSI system for the primary diagnostic use in the US. The emergence of artificial intelligence in medical diagnosis promises further growth of this field in the coming decades (3).

In a typical WSI system, we scan the tissue slide to different spatial positions and acquire the digital images using a high-resolution objective lens. The numerical aperture (NA) of the objective lens is typically larger than 0.75 and the resulting depth of field is on the micron level. The small depth of field in the WSI system poses a challenge to acquire in-focus images of the tissue sections with uneven topography. It has been shown that poor focus is the main culprit for poor image quality in WSI (4,5). The autofocusing strategy, therefore, becomes a main consideration for the image quality in the WSI system (6,7). To address the autofocusing challenge, many WSI systems create a focus map prior to the scanning process. For each focus point on the map, the system will scan the sample to different axial positions and acquire a z-stack. The z-stack images will be processed according to a figure of merit, such as Brenner gradient or entropy. The best focal position corresponds to the image with a maximum figure of merit. This process will be repeated for other tiles of the tissue slide. Surveying the focus positions for every tile requires a prohibitive amount of time. Most existing systems select a subset of tiles for focus point mapping (typically > 25). The subset of focus points can be interpolated to re-create the focus map for the entire tissue slide (6,7).

There are two major limitations for the focus map surveying method in current WSI systems. First, creating a focus map requires a significant amount of overhead time. In particular, the acquisition of a z-stack requires the sample to be static; continuous x-y motion is typically not allowed in this process. Motion acceleration and deacceleration are time-consuming for moving the sample to certain positions. Second, focus map surveying requires high positional accuracy and repeatability of the mechanical system. We need to know the absolute axial position of the sample in order to bring the sample back to the right position in the later scanning process. As a result, the cost of the WSI system is often prohibitive for many applications, such as the frozen section procedure during surgery.

To tackle the limitations of the focus map surveying method, one approach is to add additional camera(s) to perform dynamic focusing while the sample is in continuous motion (6-9). The use of additional camera(s) and the alignment to the microscope system is, however, not compatible with existing WSI platforms. Making the system more complicated also does not address the cost issue. We have recently demonstrated the use of a single camera system for rapid focus map surveying (10). This approach, however, does not address the overhead time issue and still requires high positional accuracy and repeatability of the mechanical system.

Here we report a novel WSI scheme based on a programmable LED array for sample illumination. In between two regular brightfield image acquisitions, we acquire one additional image by turning on a red and a green LED for color multiplexed illumination. We then identify the translational shift of the red- and green-channel images by maximizing the image mutual information or cross-correlation. The resulting translational shift is used for dynamic focus tracking in the scanning process. Our scheme requires no focus mapping, no secondary camera or additional optics, and allows for continuous sample motion in the focus tracking process. Since we track the differential focus positions during adjacent acquisitions, there is no positional repeatability requirement in our scheme, enabling a robust modality for building cost-effective WSI platforms. We demonstrate a prototype WSI platform with a mean focusing error of ~0.3 microns in this work. The use of programmable LED-array illumination also enables other imaging modality in the reported system, such as 3D tomographic imaging, darkfield imaging, phase contrast imaging, and Fourier ptychographic imaging (11-13).

## Methods

### *WSI System with programmable LED-array illumination*

Figure 1A-1B show the reported WSI system with a programmable LED array for sample illumination. We use a high-NA Nikon objective lens (20X, 0.75 NA) and a Nikon photographic lens (Nikon AF-S VR 105mm f/2.8G) to form a microscope system. A 20-megapixel color camera (Sony IMX 183, 2.4 μm pixel size) is used to acquire the digital images. We mount the objective lens on a motorized axial stage (ASI LX-50A) for adjusting the focus position. The sample is mounted on a motorized x-y stage (ASI MS-2000) for the scanning process. In our implementation,

the condenser lens is replaced by an 8 by 8 programmable LED array (APA102-2020 SMD LED) under the sample (11). The maximum incident angle is matched to the NA of the objective lens for regular brightfield image acquisition shown in Fig. 1C. In between two brightfield acquisitions, we turn on a red and a green LED for dynamic focus tracking. As shown in Fig. 1D, these two LED elements illuminate the sample from two opposite incident angles and the illumination NA is ~0.4 in our setup. If the sample is placed at a defocus position, there will be a translational shift between the red and the green channels from the captured color image. By identifying this translational shift, we can recover the focus position of the tissue slide. We also note that, a larger illumination NA leads to a larger image difference between the red and green channels. A smaller illumination NA, on the other hand, leads to a smaller translational shift between the red and green channels. A NA of ~0.4 is a good compromise between these two considerations

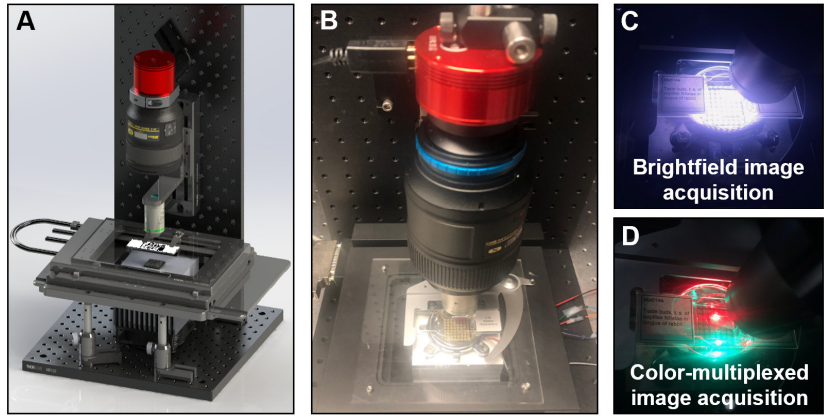

Fig. 1. The reported WSI platform with a programmable LED array for sample illumination. (A) The schematic of our platform. (B) The prototype setup. (C) Brightfield image acquisition by turning on all 64 LEDs in the array. (D) Color-multiplexed image acquisition for dynamic focus tracking. A red and green LED are used to illuminate the sample from two opposite incident angles. The translational shift between the red and the green channels from the captured color image will be used to recover the focus position during the scanning process.

The workflow of the WSI process is shown in Fig. 2. In step 1, we turn on all 64 LEDs to acquire a brightfield image of the sample. In step 2, we move the x-y stage to the next position and turn on the red and green LEDs to illuminate the sample from two opposite incident angles. In step 3, we identify the translational shift between the red and the green channels of the captured color image. In step 4, we move the z-stage to the focus position calculated by step 3. Lastly, we repeat steps 1-4 for other tiles of the tissue slide. We also summarize the timing diagram in Fig. 2, where the scanning of the x-y stage is implemented in parallel with the acquisition of the color multiplexed image. In our current prototype setup, it takes ~0.04 seconds to acquire one image, and the motion of the x-y stage takes ~0.2 seconds. As we will discuss later, we can use cross-correlation or mutual information optimization to calculate the translational shift between the red and green channels. The total time for one cycle operation is ~0.33 seconds via the cross-correlation approach and ~0.35 seconds via mutual information maximization approach in our prototype setup.

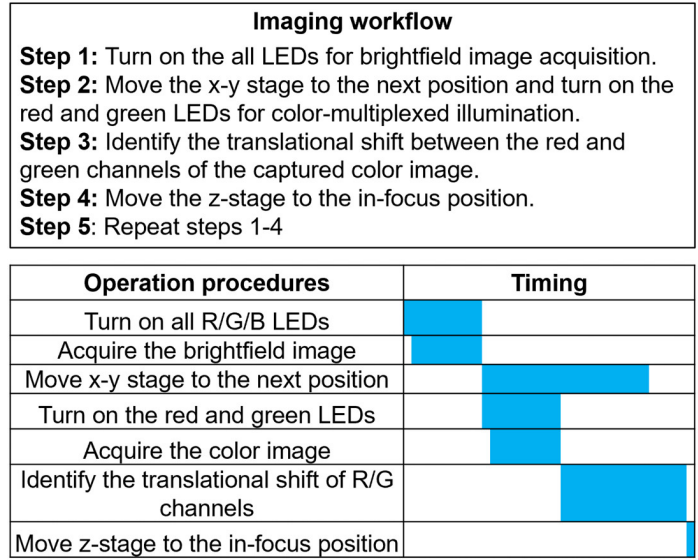

Fig. 2. Imaging workflow and the timing of different procedures.

*Translational shift between the red and green channels*

Figure 3A-3B show two captured z-stacks of a blood smear with Wright's stain and a sinistral kidney cancer section with hematoxylin and eosin stain. We can see that the separation between the red and green channels increases as the defocus distance increases. The relationship between the translational shift (in pixels) and the defocus distance is shown in Fig. 3C, where the slope is determined by the illumination angles of the red and green LEDs. If we know about the translational shift between the red and green channels, we can obtain the defocus distance based on the calibration curve in Fig. 3C.

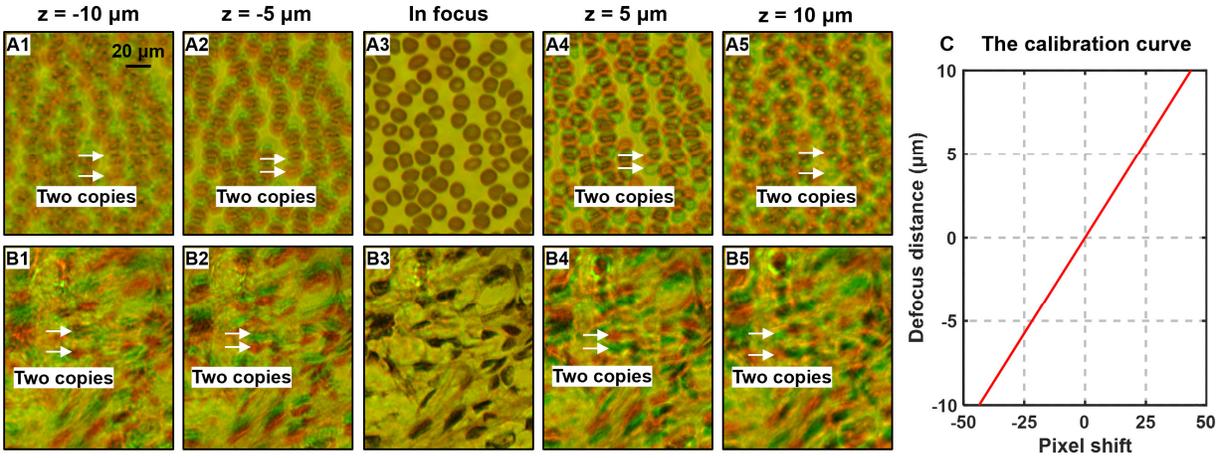

Fig. 3. The captured z-stacks under R/G multiplexed illumination for a blood smear (A) and sinistra kidney cancer section (B). (C) The relationship between the translational shift and the defocus distance.

Given a captured color image under the R/G multiplexed illumination, there are two approaches to calculate the translational shift between the two channels. The first approach is to calculate the cross-correlation between the two channels. The translational shift can be identified by locating the maximum point of the cross-correlation plot. This approach, however, assumes the two-channel images are identical to each other. If the two-channel images are substantially different from each other, the cross-relation approach would fail. The second approach is to maximize the mutual information (joint entropy) of the two channels. Mutual information is a measure of image matching, and it does not require the signal to be the same in the two images. It is a measure of how well you can predict the signal in the second image, given the signal intensity in the first. Mutual information has been widely used to match images captured under different imaging modalities (14,15). In our implementation, we use a gradient descent algorithm to maximize the mutual information of the two channels with sub-pixel accuracy. To ensure the convergence of gradient descent, we use 5 iterations in the optimization process. The performance of mutual information maximization is, in general, better than that of the cross-correlation approach. Figure 4 compares the performance between the two approaches. In this comparison, we obtain the ground truth in-focus position by maximizing the Brenner gradient (16). We then move the sample to defocus positions and quantify the focusing error using the two approaches. Different sub-sample ratios are also tested in this investigation. Mutual information maximization with a sub-sample ratio of 3 gives the most accurate result in this test. However, the processing time of the mutual-information approach is also longer than that of the cross-correlation approach. In the current implementation, we use gradient descent because of its simplicity and the calculation speed. Other advanced optimization schemes can also be used to find the translational shift between the two channels.

| Method | Subsample ratio | # of tiles | Time (s) | Focusing error (μm) |
|---|---|---|---|---|
| Cross correlation | 1 | 594 | 0.044 | 0.30 ± 0.29 |
| | 3 | 594 | 0.006 | 0.35 ± 0.29 |
| | 5 | 594 | 0.003 | 1.46 ± 0.47 |
| | 7 | 594 | 0.001 | 1.95 ± 0.57 |
| Mutual information | 1 | 594 | 0.279 | 0.29 ± 0.29 |
| | 3 | 594 | 0.065 | 0.27 ± 0.28 |
| | 5 | 594 | 0.046 | 0.54 ± 1.82 |
| | 7 | 594 | 0.041 | 3.56 ± 7.03 |

Fig. 4. Performance comparison between the cross-correlation approach and the mutual-information approach. Subsample ratio *n* means sampling the captured image every *n* pixel for calculating the two-channel shift.

*Color crosstalk correction*

Another important consideration in our approach is to correct the spectral crosstalk between the red and green channels. The color-crosstalk model can be described as:

$$I_R(x,y) = O_R(x,y) + w_{gr} \cdot O_G(x,y) \tag{1}$$

$$I_G(x,y) = w_{rg} \cdot O_R(x,y) + O_G(x,y) \tag{2}$$

where $I_R(x,y)$ and $I_G(x,y)$ are the red and green channels of the captured color image with both red and green LEDs turning on simultaneously. $O_R(x,y)$ is the red channel of the captured image under only red LED illumination. $O_G(x,y)$ is the green channel of the captured image under only green LED illumination. $w_{rg}$ and $w_{gr}$ are color-crosstalk coefficients, which can be estimated via:

$$w_{gr} \approx \frac{1}{M \cdot N} \sum_{x,y} \frac{I_R(x,y) - O_R(x,y)}{O_G(x,y)} \tag{3}$$

$$w_{rg} \approx \frac{1}{M \cdot N} \sum_{x,y} \frac{I_G(x,y) - O_G(x,y)}{O_R(x,y)} \tag{4}$$

Based on the estimated $w_{gr}$ and $w_{rg}$, the corrected red and green channels can be obtained via:

$$I_{R,corrected}(x,y) = \frac{I_R(x,y) - w_{gr} \cdot I_G(x,y)}{1 - w_{gr} \cdot w_{rg}} \tag{5}$$

$$I_{G,corrected}(x,y) = \frac{w_{rg} \cdot I_R(x,y) - I_G(x,y)}{w_{gr} \cdot w_{rg} - 1} \tag{6}$$

where $I_{R,corrected}(x,y)$ and $I_{G,corrected}(x,y)$ are the corrected red and green images. We perform an experiment to validate the performance of this color correction procedure. Figure 5A shows the ground truth of the red and green channels of a blood smear sample, which is captured under single color LED illumination. Figure 5B shows the color-multiplexed images under both red and green LED illumination. Figure 5C shows the recovered images using the correction procedure. We also plot the intensity along the solid lines in Fig. 5A-5C. We can see that the corrected intensity profile is in good agreement with the ground truth profile. In Fig. 5E, we compare the performance with and without color correction. The predicted defocus distances are in a good agreement with the calibration curve after the color crosstalk correction.

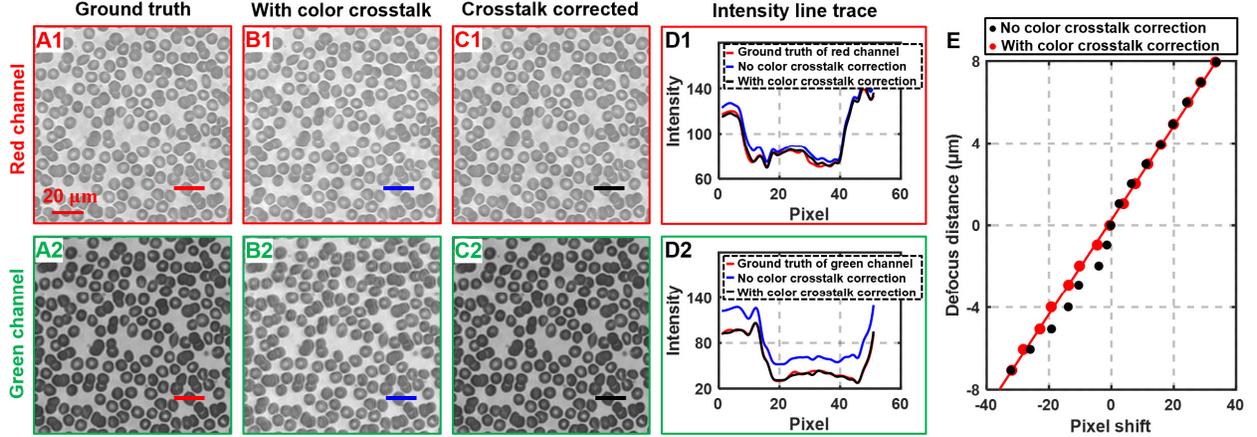

Fig. 5. Color crosstalk correction for translational shift estimation. (A) The captured ground truth images of the red and green channels under single color LED illumination. (B) The captured color-multiplexed images under both red and green LED illumination. (C) Images in red and green channels after color crosstalk correction. (D) The intensity traces along the red and green solid lines in Fig. 5(A-C). (E) The predicted defocus distances with (red dot) and without (black dot) color crosstalk correction.

**Results**

The reported WSI scheme is able to track the focus position with continuous sample motion. As shown in Fig. 1D, the red and green LEDs are aligned along the y-direction. Therefore, the translational shift between the red and green channels is along the y-direction. This allows us to introduce motion blur along the x-direction for the captured color-multiplexed images. Figure 6A shows the captured static images in the red and green channels. Figures 6B and 6C show the corresponding images with different amounts of motion blur along the x-direction. In Figs. 6D and 6E, we test the focus tracking performance with 150- and 500-pixel blur. In this experiment, we move the sample to different known defocus distances and capture the color-multiplexed images. We then calculate the translational shifts using the mutual-information approach and plot the blue dots on Figs. 6D and 6E. We can see that the reported scheme is robust against motion blur if the blur is along a direction perpendicular to the direction of the translational shift. One future direction is to analyze why the mutual information approach is effective to identify the translational shift regardless of the motion blur. The camera exposure time for color multiplexed illumination is ~15 ms in our prototype setup. The 500-pixel motion blur allows us to move the sample at the speed of 10 mm/s. A higher speed can be achieved by reducing the exposure time with a readout gain.

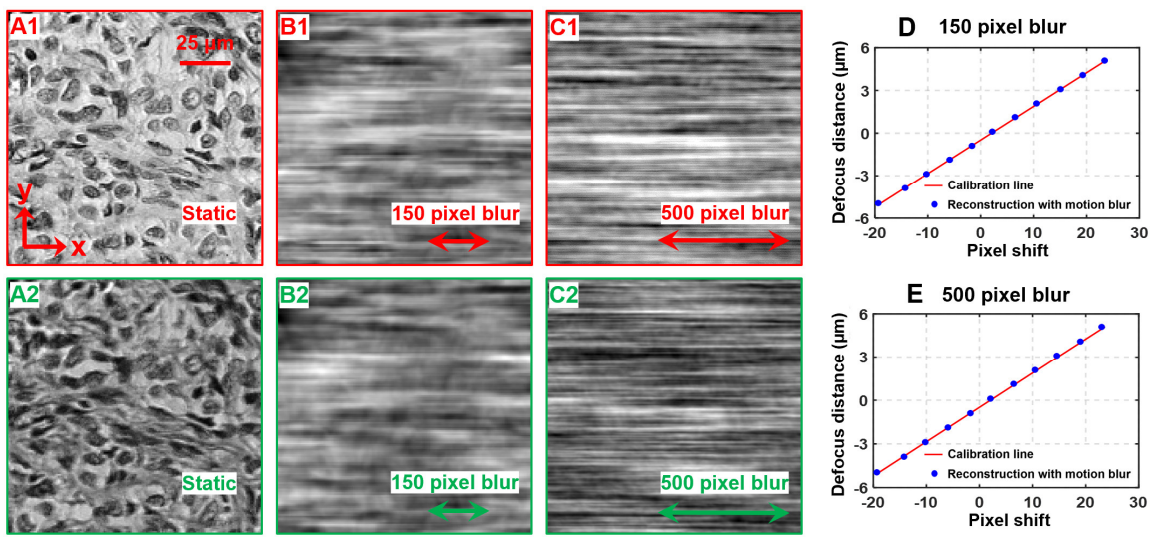

Fig. 6. Effect of motion blur. The captured red and green channels without motion blur (A), with 150-pixel motion blur (B), and with 500-pixel motion blue (C). The comparison of the calibration curve and the covered blue dots with 150-pixel motion blur (D), and 500-pixel motion blur (E).

We quantify the focusing performance in Fig. 7 for 5 different samples and 2970 tiles. The ground truth for the in-focus position is calculated based on an 11-point Brenner gradient method in an axial range of 10 µm (1 µm per step) (16). The mean focusing error is ~0.29 µm using the mutual-information approach, which is well below the ±0.7 µm depth of field range.

Based on the reported WSI scheme, we also create a differential focus map with continues sample motion in Fig. 8A (~200 pixels motion blur based on the motion speed and the exposure time of the camera). This map is different from the conventional focus map as we only track the focus difference between adjacent tiles. The corresponding high-resolution whole slide image is shown in Fig. 8B. The acquisition time for this image is ~54 seconds using mutual information maximization in the dynamic focus tracking process.

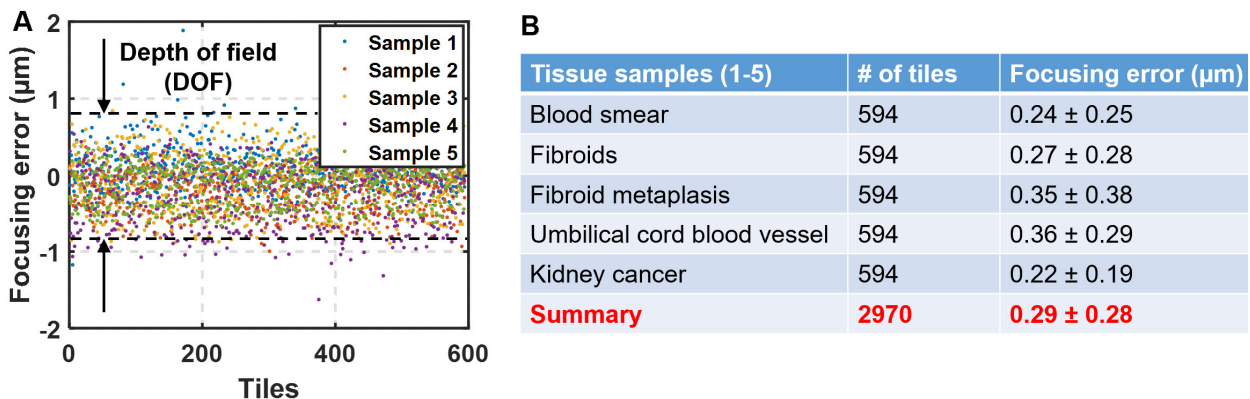

Fig. 7. Quantification of the focusing error for 2970 tiles in 5 samples. (A) The focusing errors plotted for different tiles of 5 samples. (B) Summary of the performance.

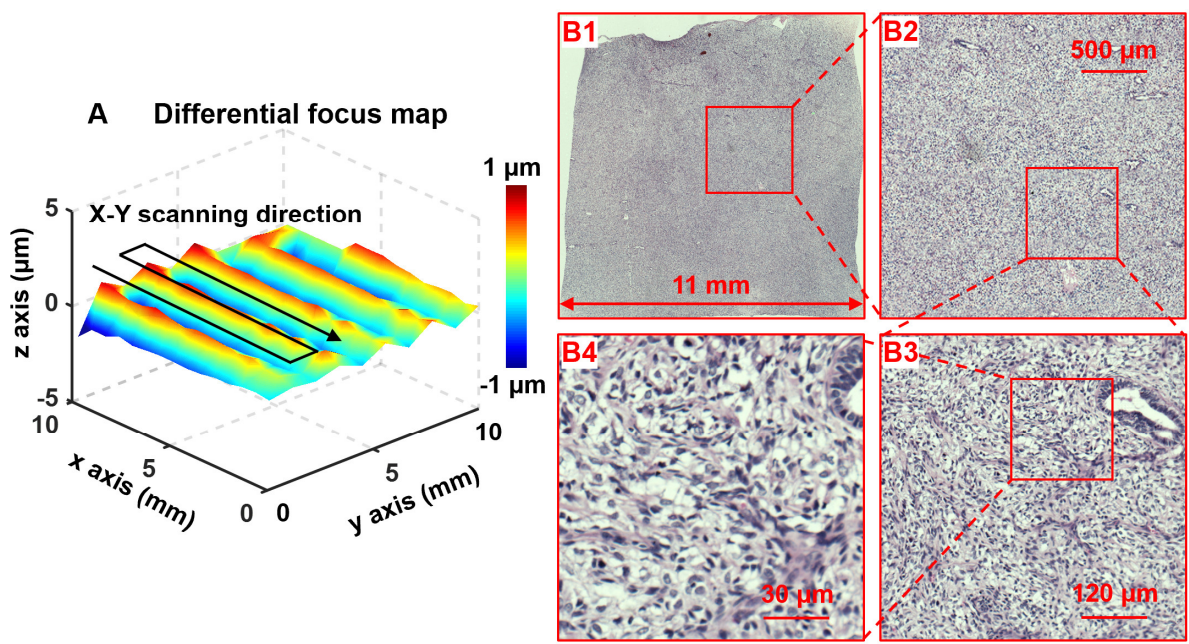

Fig. 8. (A) The generated differential focus map with continues sample motion (~200 pixels motion blur). (B) The captured whole slide image using the reported approach. The acquisition time for this image is ~54 seconds using the mutual information maximization in the dynamic focus tracking process.

## Discussion

In summary, we report a novel WSI scheme based on a programmable LED array for sample illumination. In between two regular brightfield image acquisitions, we acquire one additional image by turning on a red and a green LED for color multiplexed illumination. The translational shift between the red and green channels of the captured image is then used for dynamic focus tracking in the scanning process. We discuss two approaches to calculate the translational shift:

cross-correlation maximization and mutual information maximization. The cross-correlation approach assumes the two-channel images are identical to each other. The mutual-information approach does not require the images to be the same. It is a measure of image matching and, in general, more robust and accurate for identifying the translation shift. In a practical implementation, we can use the cross-correlation approach to get an overview of the sample property and employ the mutual-information approach to identify the translational shift. Figure 9A shows the brightfield images of a two-layer sample captured at two different axial positions. For this sample, we use color-multiplexed illumination to acquires a color image and compute the cross-correlation between the red and green channels. Figure 9B shows the cross-correlation curve, where the two peaks indicate the two layers of the sample. With this curve, we can then maximize the mutual information to locate the precise layer positions.

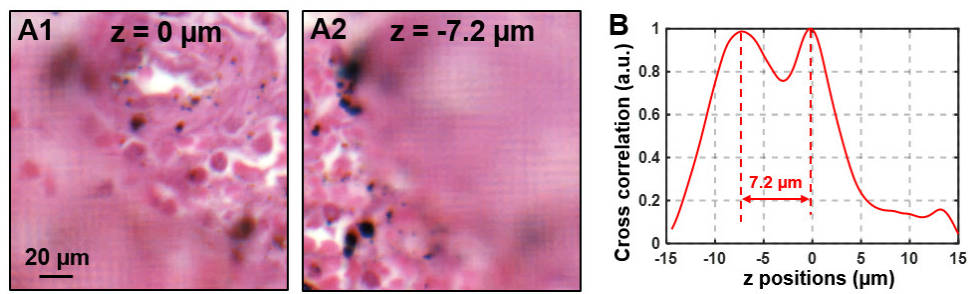

Fig. 9. (A) The captured brightfield images of a two-layer sample. The periodic pattern is due to the use of LED array illumination. For a larger defocus distance, it will generate multiple copies of the object. (B) The cross-correlation curve based on the red and green channels of the captured color image, where two peaks indicate the two layers of the sample.

Compared to previous implementations, our scheme requires no focus map surveying, no secondary camera or additional optics, and allows for continuous sample motion in the dynamic focus tracking process. It may enable the development of cost-effective WSI platforms without positional repeatability requirement. It may also provide a turnkey solution for other high-content microscopy applications.

## Acknowledgments
This work was in part supported by NSF 1555986 and NSF 1700941.